# Coherent two-dimensional multiphoton photoelectron spectroscopy of metal surfaces

Marcel Reutzel[‡], Andi Li, and Hrvoje Petek[‡]

*Department of Physics and Astronomy and Pittsburgh Quantum Institute, University of Pittsburgh,*

*Pittsburgh, Pennsylvania 15260, USA*

email: mar331@pitt.edu, petek@pitt.edu

Abstract.

Light interacting with metals elicits an ultrafast coherent many-body screening response on sub- to few-femtosecond time-scales, which makes its experimental observation challenging. Here, we describe the coherent two-dimensional (2D) multi-photon photoemission study of the Shockley surface state (SS) of Ag(111) as a benchmark for spectroscopy of the coherent nonlinear response of metals to an optical field in the perturbative regime. Employing interferometrically time-resolved multi-photon photoemission spectroscopy (ITR-mPP), we correlate the coherent polarizations and populations excited in the sample with final photoelectron distributions where the interaction terminates. By measuring the non-resonant 3- and 4-photon photoemission of the SS state, as well as its replica structures in the above-threshold photoemission (ATP), we record the coherent response of the Ag(111) surface by 2D photoemission spectroscopy and relate it to its band structure. We interpret the mPP process by an optical Bloch equation (OBE) model, which reproduces the main features of the surface state coherent polarization dynamics recorded in ITR-mPP experiments: The spectroscopic components of the 2D photoelectron spectra are shown to depend on the nonlinear orders of the coherent photoemission process $m$ as well as on the induced coherence $n$.

# I. INTRODUCTION

The quantum opto-electronic response of metals is dominantly coherent, as manifested by specular reflection of light, yet large bandwidths and ultrafast phase relaxation processes of the many-body electronic system impede its time-resolved spectroscopic investigation [1-5]. The coherent responses of metals are evident in highly nonlinear optical interactions like above-threshold multiphoton photoemission [6-10] and high-harmonic generation [11,12]. Such processes epitomize the strong-field, attosecond physics of atomic or molecular gases, as well as electron emission from nanostructured metal tips [13-16]. In lightwave electronics [17,18], exemplified by high-harmonic generation in solids, optical driving of PHz currents, and imaging of single molecule motion by THz field driven electron tunneling [19], ultrafast field driven coherent processes are opening promising topics of basic and applied research on quantum coherent processes in solid-state systems [20]. Atomic gas phase models for high-harmonic generation [18,21] have been applied to ultrafast high-field optical responses of solids, such as photon dressing of Floquet-Bloch states in topological insulators [22], laser assisted photoemission from metal surfaces [23] employing attosecond streaking [24-27], high-order mPP from sharp metal tips [14,28-30], and ultrafast coherent microscopy of plasmonic fields [31,32]. Yet many coherent processes, such as the ultrafast screening of optical fields in metals [33], are unique to the solid-state. Thus, the ability to probe and characterize how coherent fields in the solid-state [34,35] drive nonlinear phenomena can enable fundamental studies and coherent control of electronic processes and interactions [2,4], as well as applications such as generation of ultrashort electron pulses [28,36,37] for time-resolved diffraction [38] and microscopy [39,40] with potentially attosecond time resolution [34,35].

While coherence is essential for probing and harnessing of attosecond time scale high-field quantum phenomena, its effects are more easily examined at lower fields by perturbative, nonlinear mPP spectroscopy of the surface and bulk electronic bands of metals [1-4]. Detailed information on the electronic structures, as well as phase and energy relaxation dynamics at surfaces, interfaces, and complex adsorbate structures, has been obtained by time-resolved multi-photon photoemission (TR-mPP) spectroscopy [41-47]. In mPP, the combined energy of two or several photons excites electrons from the occupied bands below the Fermi level ($E_F$) to the photoemission continuum above the vacuum level ($E_{vac}$), via real or virtual unoccupied intermediate states. For mPP, with $m \geq 3$, high optical field strengths are necessary to detect sufficient photoelectron yields, while taking care that sample damage or space charge do not compromise the spectroscopic measurement. In addition to the lowest-order mPP signals, when exciting with ultrashort laser pulses having perturbative optical field strengths, higher-order nonlinear effects, like above-threshold photoemission (ATP) [6-10], can also be detected. ATP is a defining feature of attosecond physics, where the strong optical fields drive coherent electron motion to above the threshold energy for photoemission, yet it has not been critically examined in the perturbative regime.

Interferometric time-resolved multi-photon photoemission (ITR-mPP) is an effective multidimensional spectroscopy for evaluating quantum coherence in solids and solid surfaces; it has been developed and applied to study of dephasing of surface and bulk bands in metals [1,48,49], the coherence in interfacial charge transfer processes [50,51], and the formation of transient excitons at Ag(111) surface [33,52-54]. Cui *et al.* [33] developed and applied the multidimensional capabilities of ITR-mPP [41,55] to record movies of the coherent transformation of the transient exciton formed by resonant two-photon excitation of the $n = 1$ image potential (IP) state ← Shockley surface (SS) state transition of Ag(111) by collecting photoelectron energy ($E$), and parallel momentum ($k_{\parallel}$) distributions with an imaging electron spectrometer, while scanning the optical delay time ($t$) between identical pump and probe pulses with ~50 as (sub-optical cycle) resolution.

Optical phase-resolved $E(k_{\parallel},t)$ movies recorded by ITR-mPP enable acquisition and analysis of the coherent electron dynamics in the time, frequency, energy, and momentum domains. Fourier transforms (FTs) of delay dependent photoelectron energy distributions for a given $k_{\parallel}$ give two-dimensional (2D-FT) photoelectron spectra, which correlate the coherent polarization frequencies (energies) induced in the sample with the photoelectron energies that are detected. Such 2D-FT spectra record the linear and nonlinear current-current correlations, and thus enable analysis of coherences that produce the mPP signal [56-59]. Whereas optical spectroscopy measures energy and momentum integrated optical transitions between states, the merit of mPP spectroscopy is that it measures the $E$ and $k_{\parallel}$ resolved excitations that terminate at a specific photoelectron energy and momentum. Because $E$ and $k_{\parallel}$ are conserved, the mPP spectra of surface bands are homogeneously broadened. Therefore, by contrast to the 2D optical spectroscopy, which educes the coherent optical response from correlations between tunable excitation and detection fields [60,61], an ITR-mPP experiment correlates the coherent polarizations and populations with the final photoelectron distribution [33]. The coherent nonlinear polarizations recorded by ITR-mPP, are directly related to the high-optical-harmonic generation at metal surfaces [58,62]. In the perturbative regime, a field $E$ induces a polarization $P$ that can be described by $P = \sum_{l=1}^{\infty} l\epsilon_0 \chi_l E^l$, where $l$ is an integer and the susceptibility $\chi_l$ describes the energy, momentum, and time integrated linear and nonlinear system responses to the applied field [58]. In ITR-mPP experiments on Ag(111) surface to be described, we detect the coherent polarizations, which derive from the dominant energy and momentum resolved contributions to the nonlinear response up to $l = 10$.

Coherent 2D-FT photoelectron spectroscopy has been applied to several samples [33,51,63]; coherent excitation pathways that participate in the mPP process have been recorded. Here, we examine the signatures of coherence in 2D-FT photoelectron spectra for nonresonant 3PP, 4PP and 5PP, including the ATP excitation, of the partially occupied SS at the pristine, noble metal Ag(111) surface. The well-known surface electronic structure of Ag(111) [Fig. 1(a)] facilitates the spectroscopic assignments of mPP and simulation of the coherent polarization dynamics by an optical Bloch equations (OBE) model [58,64-66]. The simulations reproduce the main features of the experimental 2D-FT spectra, providing insights into the coherent multi-photon excitation as well as ATP at metal surfaces.

## II. METHODS

### A. Experimental setup: ITR-mPP

Interferometrically time-resolved mPP experiments [33,41,55] are performed in an ultrahigh vacuum chamber with a base pressure of <$10^{-10}$ mbar. The Ag(111) surface is prepared by Ar$^+$ ion sputtering (20 min, 1500 V, 3 µA) and subsequent annealing (10 min, 550 K); its quality is judged by a sharp SS peak in the mPP spectra. The sample is cooled to ~90 K for mPP experiments to minimize dephasing by electron-phonon interaction.

mPP spectra are excited with the output of a noncollinear optical parametric amplifier (NOPA) pumped by a Clark MXR Impulse Fiber laser oscillator-amplifier system operating at a 1

MHz pulse repetition rate [67]. The fundamental and second harmonic outputs of the NOPA provide tunable ultrafast laser pulses in a 930 - 270 nm range with <20 fs pulse durations. 3PP and 4PP are excited by focusing near-IR pulses onto the sample at an average power of ~70 mW. The Keldysh parameter for the excitation, $\gamma \approx 20$, is much above $\gamma < 1$ where the nonperturbative effects such as field-induced tunneling begin to dominate [16]; thus, our experiments are in the perturbative regime where the mPP spectra are generated by dipole coupling of the surface and bulk bands [20,68]. 2PP spectra with UV light are excited with an average power of ~5 mW. After compensation with negative dispersion mirrors, the NOPA pulses are characterized by recording ITR-mPP autocorrelations on the polycrystalline Ta sample holder at the location of the Ag(111) sample [63].

For the interferometric measurements, identical pump-probe pulse pairs are produced with a Mach-Zehnder interferometer and their delay is scanned by piezoelectrically varying the path-length of one arm [1]. One output of the interferometer is directed to the sample as the mPP excitation source, while the other is transmitted through a monochromator to generate interference fringes that calibrate the time delay [1,33,41]. A hemispherical electron energy analyzer and a position sensitive electron counting detector produce energy and angle resolved 2D spectra. A series of spectra are taken as a function of pump-probe delay in ~100 as intervals to generate 3D $E(k_{\parallel},t)$ movies. Multiple pump-probe scans are averaged to improve the counting statistics [33].

### B. OBE Simulation

We simulate the coherent ITR-mPP data with OBE simulations based on a ($m$+1)-level excitation schemes ($m$: number of absorbed photons); the general case for a 3-level system is shown in Fig. 1(b). The OBE provide a semi-classical description of optical excitation of atomic or molecular systems [64], and thus, can only include heuristically the many-body effects associated with the quasiparticle formation and dephasing. For homogeneously broadened bands of a metal surface, the mPP process involving *m*-photon absorption can be described by an (*m*+1)-level system, as has been verified previously [1-3,41,65,66,69,70]. Time evolution of the (*m*+1)-level system is modeled by a density matrix $\hat{\rho}$, where the diagonal elements $\hat{\rho}_{ii}$ represent populations of the (*m*+1)-levels, and the off-diagonal elements $\hat{\rho}_{ij}$ describe the coherences or polarizations generated by the optical field [cf. Fig. 1(b)]. In a nonresonant mPP process the photoinduced coherences transfer populations from the initial to the final state, as described by The Liouville-von Neumann equation

$$i\hbar \frac{\partial \hat{\rho}}{\partial t} = \left[\hat{H}, \hat{\rho}\right] + \frac{\partial \hat{\rho}_D}{\partial t}.$$

$\hat{H}_0$ and $\hat{H}_{LM}$ in the full Hamiltonian $\hat{H} = \hat{H}_0 + \hat{H}_{LM}$ describe the unperturbed system, and semi-classically treated light-matter interaction in the dipole approximation, respectively. The dissipation matrix $\hat{\rho}_D$ phenomenologically describes the damping of the density matrix terms [65,66], where the diagonal $\hat{\rho}_{Dii}$ and off-diagonal $\hat{\rho}_{Dij}$ elements represent the inelastic decay rates $1/T_{1i}$ of level $i$ and dephasing of the coherences $1/T_{2ij}^*$ between the levels $i$ and $j$, respectively.

In the OBE calculations, we have chosen the parameters of $\hat{\rho}$ and $\hat{\rho}_D$ as follows: The energy levels and the photon energy are defined by the experiment; the laser pulse width is set to 20 fs. The initial and the final state inelastic decay times $T_{1i}$ are set to infinity ($T_{1SS} = T_{1F} \gg 1$ fs) [66], while that for the *n* = 1 IP state is its measured value of $T_{1IP}$ = 30 fs [71]. For *m* ≥ 3 (4- and 5-level system), the ladder climbing is described by virtual intermediate states (VI), for which the inelastic decay times are set to $T_{1VI} \gg 1$ fs. The dephasing times $T_{2ij}$ are set in a range between 15 and 1500 fs. We emphasize that we only aim to qualitatively simulate the electron dynamics.

# III. RESULTS AND DISCUSSION

The occupied electronic structure of the Ag(111) is known from one-photon angle-resolved photoemission spectroscopy [72-75]. The unoccupied band structure has been studied by inverse photoemission [76] as well as 2PP and 3PP spectroscopy [71,77-84]. The focus of these mPP experiments has been the image potential state series ($n$ = 1, 2, …), their excitation from the occupied SS state including formation of an excitonic state [33], their lifetimes, and the bulk transition from the lower, $L_{sp}$, to the upper, $U_{sp}$, sp-band [33,71,77-84].

In Fig. 2, we show energy- and $k_\parallel$-resolved 2PP spectra taken by excitation with $\hbar\omega_l$ = 3.32 eV photons. 2PP spectroscopy of Ag(111) in this energy range is well documented in literature; our binding energies and $k_\parallel$-dispersions are in agreement with these reports [81,82]. The 2PP spectra are dominated by the SS state and the sp-band, which are, respectively, detected by nonresonant and resonant ($L_{sp}$ to $U_{sp}$) coherent two-photon absorption; the SS state is occupied in a narrow $k_\parallel$-range where it exists below $E_F$ [excitation diagram in Fig. 1(a)] [82]. The $n$ = 1 IP state is detuned from resonance with SS, and hence, has a low intensity. Remarkably, the IP state is recorded only in the $k_\parallel$-range where the SS state is occupied (compare the blue and grey line-profiles in Fig. 2). We thus conclude that the IP state is populated by nonresonant one-photon excitation from SS through ultrafast dephasing [85], rather than by resonant absorption (two-photon) from the bulk sp-band, which is also possible, but would not produce the distinct $k_\parallel$-range of SS.

## A. OBE simulation

Because the OBE approach has proven useful to interpret, simulate, and analyze ITR-mPP data [1,41,50,51,63,86], in this section we present an OBE simulation of the coherent electronic response of Ag(111) surface, by calculating the 2PP signal as the delay is varied interferometrically between identical pump and probe pulses [Fig. 3(a)]. The calculation assumes the 3-level scheme in Fig. 3(b), which qualitatively models the 2PP process of Fig. 2. Subsequently, we address the related coherent excitation dynamics of the SS state of Ag(111) in higher-orders of multi-photon excitation by presenting the ITR-mPP [$m$ = 3, 4, 5 (ATP)] data and their OBE simulations. The surface and bulk states appear independently in the mPP spectra, and hence we do not model the $L_{sp}$-$U_{sp}$ transition, which has been described previously [81,82].

In the OBE calculation, the SS state is excited into the final state continuum by coherent two-photon absorption in near resonance with the IP state [Fig. 3(a)]. As the interaction proceeds, the photoexcited bandwidth is wide and encompasses the IP←SS transition, but with time it narrows through destructive interferences; some IP state population is created by dephasing of the IP-SS coherence. Thus, the final state population includes both the coherent two-photon absorption from SS state as well as photoemission of the dephased population in the IP state. The photon energy is such, that detuning of the IP ← SS resonance is sufficiently large to separate the spectral features of the (1+1)-photon process via the IP state from the coherent 2-photon absorption from the SS state; quantum interference caused by excitation of final states through two differing pathways is negligible [50]. Thus, the calculated 2PP signal is predominantly produced by rectification of two-photon coherence generated by excitation of the occupied SS level.

In Fig. 3(a) (right), we show cross sections [interferometric two-pulse correlations (I2PCs)] through the calculated two-pulse interferogram, for the final state energies of the SS and IP states; similar correlation traces oscillating at the driving laser frequency and its second harmonic were observed in the original ITR-2PP study of the SS state on Cu(111) [1]. The I2PC signal of the IP state has a slow component due to its incoherent population decay, as is measured in two-color 2PP [71]. The inset of Fig. 3(a) shows enlarged SS component of the ITR-2PP data for time delays that are comparable or longer than the laser pulse: The signal is composed of dominant coherent oscillations at the driving laser frequency ($1\omega_l$) and subordinate ones at its second-harmonic ($2\omega_l$) [1,41]. Here, we stress the ability of OBE simulations to reproduce the energy- and phase-resolved 2PP measurements of the coherent electron dynamics in Ag(111), as will be apparent in experimental results for higher order processes. Specifically, we note that interference fringes tilt towards $t = 0$ fs with increasing final state energy [inset of Fig. 3(a)], which is the *prima facie* evidence for coherence in a 2PP process.

In further analysis of the calculated ITR-mPP data, we Fourier transform it to generate 2D-FT photoelectron spectra to correlate the polarization frequencies (energies) that are excited in the sample, and appear as coherent oscillations, with final state energies where the mPP process terminates [Fig. 3(c)]. Local FT amplitude maxima of the 2D-FT spectra occur for $0\omega_l$, $1\omega_l$, and $2\omega_l$, representing the linear and nonlinear polarization of surface electrons of Ag(111) to the driving field. For a three-level system, $1\omega_l$ and $2\omega_l$ coherences have dominant contributions between the nearest neighboring $\hat{\rho}_{i(i+1)}$ and next-nearest neighboring $\hat{\rho}_{i(i+2)}$ states, respectively [Fig. 1(b)]; the $0\omega_l$, however, represents the envelope function of the interferogram, including the relatively slow incoherent population dynamics of the intermediate state [41,86,87].

Figure 4(a) shows enlarged elements of the 2D-FT photoelectron spectra of the SS state in Fig. 3(c). The polarizations that oscillate at the $0\omega_l$, $1\omega_l$ and $2\omega_l$ frequencies are rectified to produce the final states; the polarization ranges cover the widths of the driving frequency $\omega_l$ and its harmonics. The FT amplitudes of the linear ($1\omega_l$) and the second-order non-linear ($2\omega_l$) polarizations tilt with the final state energy: the coherent excitation of a discrete initial state with an ultrafast laser pulse with a broad frequency spectrum induces primarily linear polarization oscillations over a broad frequency range centered on $\omega_l$. Likewise, the nonlinear polarizations that are also excited and oscillate at $2\omega_l$ and higher harmonics. As sketched in Fig. 4(b), the spectral components of polarizations at $1\omega_l$ or $2\omega_l$ correlate with specific final states to satisfy the energy conservation and energy-time uncertainty. Thus, in a coherent *m*-photon excitation process of a discrete state, slopes of the 2D spectral features are given by $n/m$, where *m* is the order of photoemission (e.g., *m* = 2, 3, 4, …) and *n* is the order of the (non-)linear polarization ($n\omega_l$; *n* = 0, 1, 2, …, $n \leq m$) contributing to the signal. For example, for the coherent 2PP process of the SS state of Ag(111), as described by the three-level system (Fig. 3 and Fig. 4), the slopes of the $0\omega_l$, $1\omega_l$ and $2\omega_l$ FT amplitudes are $n/m$ = 0/2, 1/2 and 2/2, respectively [Fig. 4(a)].

The energy tilting of components of 2D-FT photoelectron spectra is a consequence of the tilting of interference fringes in the time-domain ITR-mPP spectra. Specifically, the tilted interference fringes in the enlarged part of Fig. 3(a) signify that the field of each frequency component of the broadband excitation laser pulse causes the induced polarization to oscillate at its frequency. Thus, if each frequency component correlates with a specific final state energy, the plot of the final state energy *vs*. the delay time will have tilted interference fringes, as observed; the tilt of the fringes gives insight into coherences induced by the optical field in the sample [51]. Similar fringe tilting is also observed in optical spectroscopies that involve interferometric scanning and frequency resolved optical spectroscopic detection of coherent responses, such as frequency-resolved optical gating or second harmonic generation [88-90].

In photoemission spectroscopy, however, it is possible that the correlation between the polarization frequency and the final state energy is erased. This happens when the inhomogeneous broadening is large, because populations at different energies oscillating at different frequencies can be excited to the same final state. Likewise, if dephasing is very fast, the polarization frequencies are poorly defined through energy-time uncertainty, which broadens the FT spectral components. In such cases, the signal oscillates with amplitude of the total electric field, rather than its frequency components, because final state energies do not correlate with specific coherences excited in the sample. Consequently, the ITR-mPP fringes do not tilt as is observed, for example, for inhomogeneously broadened polycrystalline samples.

Next, we extend our discussion to experimentally observed and OBE simulated coherences in ITR-mPP ($m$ = 3, 4, 5) measurements including their ATP components and evaluate them based on the insights from the calculated ITR-2PP data in Fig. 3. By characterizing the interferences in the time- and frequency domains, we present a systematic picture of the nonresonant mPP of the SS state as a function of photoemission and polarization order, as well as ATP in the perturbative regime.

### B. 3PP and 4PP spectroscopy of Ag(111)

We measure 3PP and 4PP spectra of Ag(111) for photon energies of $\hbar\omega_l$ = 1.40 and 1.60 eV (Fig. 5); in both cases, the SS state is prominent though excited nonresonantly as indicated by excitation diagrams in Fig. 1(a). For $\hbar\omega_l$ = 1.60 eV [Fig. 5(a)], photoelectrons are excited from the SS state by a coherent three-photon process. At a final state energy ≈1.6 eV above the 3PP spectrum of SS, its replica spectrum with the same $k_\parallel$-dispersion, but with >100 times weaker intensity, is observed. We assign this structure to 4PP, or the ATP, of the SS state. In addition, we observe similarly weak 4PP signals from the $n = 1$ IP state and bulk excitation between the $L_{sp}$ and $U_{sp}$. The energies and $k_\parallel$-dispersions of the surface states are consistent with the 3PP and 4PP processes, as well as with the 2PP data in Fig. 2 and related 2PP literature [33,71,77,82,84].

For $\hbar\omega_l = 1.40$ eV [Fig. 5(b)], 4PP spectra of the SS, the $n = 1$ IP state, and the bulk excitation between the $L_{sp}$ and $U_{sp}$ are observed. Again, replicas of the SS, the IP states, as well as of the sp-band are detected at approximately ≈1.4 eV higher energy, which we also attribute to ATP, i.e. 5PP spectra of the respective states. The $k_\parallel$-resolved ATP data has comparable dispersions as the lower-order processes and thus we anticipate that it involves coherent excitation processes. Significantly, the intensity ratios of the SS and IP states are inverted between the 4PP and ATP processes: In ATP, the highly non-linear 5-photon absorption process from the SS state seems to be more efficient than the (3+2)-photon process via the transiently occupied intermediate IP state.

The nonresonant excitation of Ag(111) with near-infrared photons is thus an excellent system to study the ultrafast coherent electron dynamics in higher-order mPP and ATP processes; we now present the corresponding ITR-mPP experiments ($m = 3, 4, 5$).

### C. Coherent 3PP and 4PP dynamics of the SS state

The nonresonant excitation of the SS state of Ag(111) provides a simple system to study high-order, perturbative, coherent electron dynamics. In Fig. 6, we show the experimental ITR-4PP data taken with the same experimental conditions as used for the 4PP spectrum in Fig. 5(b) ($\hbar\omega_l =$ 1.40 eV, $k_\parallel = 0$). The I2PC scans at the SS and IP state maxima in Fig. 6 exemplify the coherent dynamics in 4PP; the large peak-to-baseline ratios result from the high nonlinearity of the excitation process. The simultaneous measurement of mPP over a wide range of final state energies as a function of interferometric pump-probe time delay gives further insights into coherence of the photoemission process: As in the calculated ITR-2PP data in Fig. 3, tilting of the interference fringes towards $t = 0$ fs with increasing final state energy is observed (inset of Fig. 6). The tilt angles of the interference fringes correlate the induced polarization energy with the detected final state energy.

To analyze the coherent dynamics in more detail, in Fig. 7(a) we plot the SS state components of the 2D-FT photoelectron spectra obtained from the data in Fig. 6. The FT amplitudes for the $0\omega_l$-$4\omega_l$ components again tilt, respectively, with slopes of 0, 1/4, 2/4 and 3/4 ($4\omega_l$ signal is too weak to evaluate, and probably requires higher time resolution interferometric scanning to define). By applying the same arguments as for the 2PP OBE simulation, we explain the slopes with the ratio $n/m$ where now the order of the photoemission process is $m = 4$ and of the polarization is an integer multiple of the excitation frequency ($n = 0, 1, 2, 3$). In other words, to achieve 4PP from the SS state, the coherent signal is driven at harmonics of the driving frequency (numerator $n$), and different frequency components of the photoelectron energy correlate with the order of the multi-photon process (denominator $m$). To further test this interpretation, we performed an OBE simulation of the 4PP process for a 5-level system, which gives comparable 2D-FT spectral components as the experiment [Fig. 7(b)]. The larger widths of the simulated 2D-FT spectra, as compared with the experiment, reflect faster dephasing $1/T_{2ij}^*$ parameters used in the OBE simulation.

To address the effect of the order $m$ in mPP, we show 2D-FT data for ITR-3PP detection of the SS state in the supplemental material (energy level diagram in Fig. 1(a); 3PP spectrum in Fig.

5(a) for $\hbar\omega_l = 1.60$ eV, $k_\parallel = 0$). The ITR-3PP experiments as well as the corresponding 4-level OBE simulation confirm again the $n/m$ dependence of tilting of the 2D-FT spectra in a nonresonant, coherent process. The 2D-FT photoelectron spectra of the SS state thus systematically depend on the induced polarization signal ($n \leq m$) and order of the multi-photon excitation process ($m = 3, 4$).

### D. Coherent electron dynamics leading to ATP

The spectra in Fig. 5 show clear contributions of nonresonant photoemission from the SS state following $m$- and ($m$+1)-photon photoexcitation that produce the mPP and ATP signals, respectively; the two signal components are recorded simultaneously in an ITR-mPP scan in Fig. 6 and thus can be compared, even though they differ by a factor of ~100 in intensity. Here, we present the first time-resolved measurement of ATP in a solid-state system, which defines how it occurs. In Fig. 8(a), we show the ITR-mPP data for the ATP processes from both the $n = 1$ IP and SS states; the I2PC scan at the SS maximum shows a large peak-to-baseline ratio (larger than the IP state) as expected for a $5^{th}$ order non-linear processes. Again, we highlight coherence in the mPP process of the SS state by showing in the inset in Fig. 8(a) that the interference fringes tilt towards $t = 0$ fs with increasing final state energy. The time-domain experiment thus shows that 5-photon excitation (ATP) of the SS state is a coherent process.

To characterize this coherence, we generate 2D-FT photoelectron spectra of the ATP component of the SS state [Fig. 8(b)]. FT amplitudes are well structured for $0\omega_l$ (not shown) to $3\omega_l$, but higher-order components have insufficient statistics for analysis. The FT amplitudes show characteristic tilting, like that observed in 4-photon excitation in Fig. 7, which in analogy we analyze in Fig. 8(b) with the slope $n/m$, where $m = 5$, and $n = 1, 2$, and 3 for the $1\omega_l$, $2\omega_l$ and $3\omega_l$ components

The ITR-mPP data show tilted interference fringes and FT amplitudes in the time- and frequency-domains, where it is evident that the dominant mPP and the accompanying ($m$+1)-ATP processes arise from the same physics. Although ATP is usually described as occurring by generation of a photoelectron through mPP, followed by the photoelectron absorbing another photon before emerging from the surface region, our measurements show that such process does not dominate. In the perturbative regime, the optical field drives the SS state electrons to oscillate nonlinearly at $m\omega_l$ and ($m$+1)$\omega_l$ frequencies. These coherent polarizations can produce coherent light emission at the $m$ and ($m$+1) harmonics of the driving field [58], as has been observed from Cu(111) for $m = 2$ at lower fluences [91]. In addition, the mPP and ($m$+1)PP ATP signals can be produced by rectification of components of the induced polarization into photoelectron currents at the $m$ and ($m$+1) final state energies; our data show this to be the more probable scenario than the $m$-photon final state absorbing an additional photon before departing from the surface. In the case of ATP with $\hbar\omega_l = 1.40$ eV, this corresponds to a $\chi_{10}$ process. Of course, in a 5-photon process there are many other excitation pathways to reach the final state, but only the rectification of the ($m$+1) polarization would produce the observed FT amplitude slopes in Fig. 8(b). That the ATP photoelectron is generated preferentially by rectification on ($m$+1) polarization can explain why the intensities of SS and IP state are reversed between the mPP and ATP orders in Fig. 5(b). Whereas the ATP signal from the SS state results from rectification of the ($m$+1) polarization, that for the IP state first populates the intermediate state by the rectification of the $3\omega_l$ polarization followed by rectification of the $2\omega_l$ polarization to generate the ATP signal. Because susceptibilities are time-dependent [58], the cascaded process of lower order coherent responses is probably less efficient than the direct higher order process. Although both are driven by an intense electric field, the ATP processes reported here should be contrasted with the nonperturbative regime, where it involves the field-induced recollision of electrons with their atomic cores [92]. We again emphasize that under our experimental conditions, we observe no spectroscopic evidence of field-induced nonperturbative dynamics.

## IV. CONCLUSIONS

We have elaborated the coherent multi-photon photoemission of the Shockley surface state electrons from Ag(111) surface by means of ITR-mPP experiments and OBE simulations for nonlinear orders of $m$ = 2 - 5. The nonresonant energy level climbing of SS electrons in the perturbative regime is dominantly caused by oscillations of the nonlinear polarization at high orders of the driving field, as recorded by ITR-mPP measurements and their Fourier analysis; the coherence can dephase to populate nonresonant intermediate states, such as the IP states, it can be rectified to produce the mPP signal or, in higher order, its ATP replicas, or it can decay by producing harmonic emission. In time-domain measurement, the coherent electron responses of a metal surface state is evident in tilting of interference fringes because the final state energies correlate with the induced polarization frequencies. In frequency-domain, the same physics causes the Fourier amplitudes to tilt such that their slopes are given by the ratio $n/m$, where at $n$ is the (non-)linear order of the polarization and $m$ is the order of mPP process. This stands in contrast with the nonperturbative regime, where the optical field transiently distorts the atomic Coulomb potential and thereby drives electron trajectories through resonances for high harmonic emission.

Our study gives systematic insights into shapes of 2D photoelectron spectra as obtained by Fourier analysis of ITR-mPP experiments: By addressing the simplest case of the nonresonant excitation of the discrete Shockley surface state of Ag(111) we establish a benchmark for deeper understanding of more complex excitation schemes such as in the two-photon IP ← SS resonant excitation on Ag(111) [33] or proposed for multiexciton generation in organic semiconductors [93]. The coherent excitation of surface states of metals provides a simple electronic structure, which can be easily resolved and selectively excited in a mPP process. With a high-power tunable laser source, the resonant or nonresonant excitations pathways can be selected, excited, and compared. Such coherent excitation processes can occur in other solid state systems where similar experimental procedures and analysis can be applied. Coherent 2D optical spectroscopy has already been applied to excitonic correlations in GaAs [94]. The method reported here provides an alternative approach that probes the induced coherence over a wide energy range, and that is likely to be effective in strongly interacting solid state systems, that may be nonemissive.

We have also reported the first coherent spectroscopic investigation of perturbative ATP processes in solids. Our results show that ATP involving nonresonant excitations can be understood as a higher order driving of the coherent of SS electron oscillations, when compared to the main mPP signal, rather than the conventionally proposed two-step process where a photoelectron is generated and absorbs another photon before emerging into the vacuum. The energy, $k_{||}$, and time resolved ITR-mPP measurements provide insights into coherent optical processes at metal surfaces in the perturbative interaction regime.

**ACKNOWLEDGEMENTS**

The authors gratefully acknowledge support from DOE-BES Division of Chemical Sciences, Geosciences, and Biosciences Grant No. DE-SC0002313, and M. R. acknowledges support from the Alexander von Humboldt Foundation within the Feodor Lynen Fellowship Program.

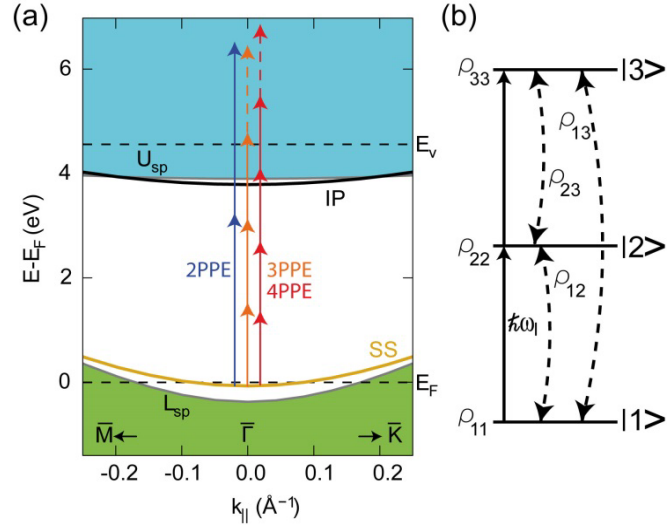

**Figure 1** | (Color online) (a) Surface projected band structure of Ag(111) after Ref. [76]. The lower ($L_{sp}$) and the upper ($U_{sp}$) bulk sp-bands are shown by green and light blue shading, respectively. The Shockley surface state (SS), which is occupied for a narrow $k_\parallel$-range, is plotted in ocher and the $n = 1$ image potential state (IP) in black. Photoemission of electrons from SS state is induced in second, third and fourth order mPP ($m$ = 2, 3, 4) processes, depending on the photon energy, as indicated by the arrows. Above threshold photoemission (ATP) in $m$ = (3+1)- and $m$ = (4+1)-photon photoexcitation is indicated by the dashed arrows. (b) Diagram of a canonical 3-level system used in OBE calculations consisting of the occupied initial state $|1\rangle$, the unoccupied intermediate $|2\rangle$ and the final photoelectron state $|3\rangle$. Upward and dashed arrows indicate population transfers and coherences between the levels, respectively, which are labeled with the corresponding elements of the density matrix $\hat{\rho}$; the state energies depend on a specific electronic structure.

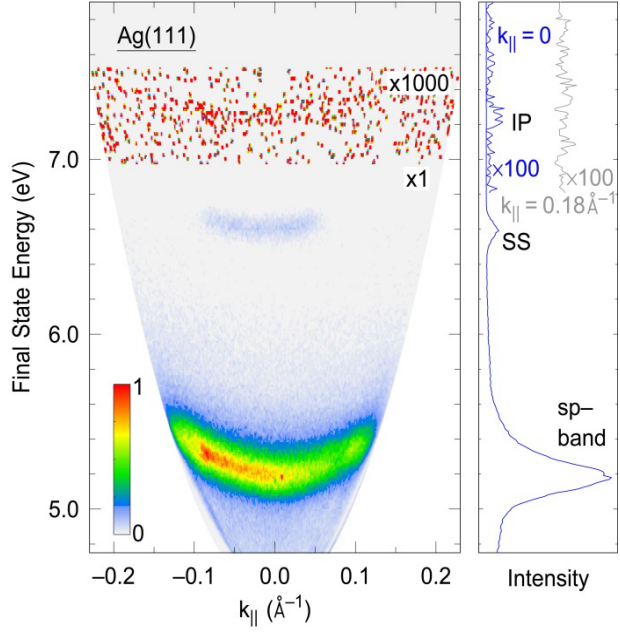

**Figure 2** | (Color online) Energy- and $k_\parallel$-resolved 2PP spectra of Ag(111) [$\hbar\omega_l$ = 3.32 eV; the corresponding excitation diagram for the SS state is given in Fig. 1(a)]; the photoemission spectral elements are labelled in the line profile (right, blue) taken at $k_\parallel = 0$. The SS state and the sp-band are detected through coherent two-photon excitation and resonant two-photon absorption from $L_{sp}$ to $U_{sp}$, respectively. The color intensity scales represent the photoelectron counts; above 7 eV, the color-scale is amplified by a factor 1000 to show the weak contribution of the $n = 1$ IP state. As emphasized by the line profiles shown for $k_\parallel = 0$ Å$^{-1}$ (blue) and $k_\parallel = 0.18$ Å$^{-1}$ (grey), the IP state is only resolved for the $k_\parallel$-range in which the SS state is occupied.

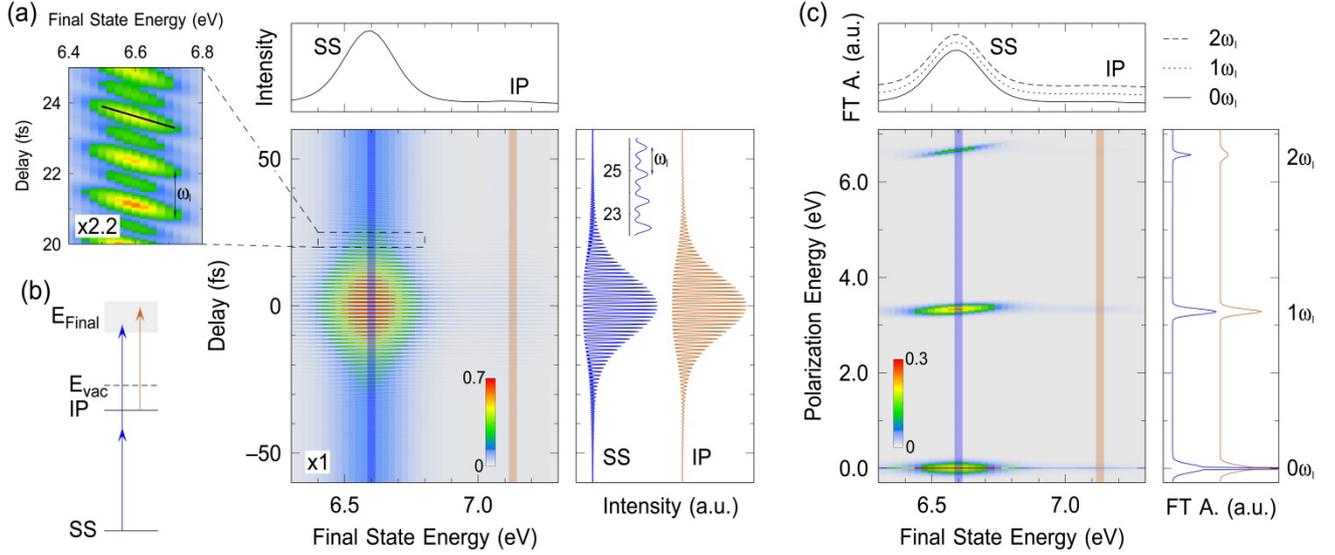

**Figure 3** | (Color online) (a) OBE simulation of an energy- and phase-resolved ITR-2PP measurement for the 3-level system shown in (b), which consists of the occupied initial state (SS), unoccupied intermediate state (IP) and final state continuum ($E_{Final}$) above the vacuum level $E_{vac}$ (grey box). The center plot in (a) shows the calculated ITR-2PP measurement, specifically population of the final state (color-coded) as a function of final state energy and pump-probe delay. Top: The calculated final state energy spectrum consisting of the SS and IP states is a profile through the ITR-2PP data at $t = 0$ fs. Right: Interferometric two-pulse correlation traces are profiles through ITR-2PP for the energies of SS and IP states; oscillations caused by interferences at the driving laser frequency $1\omega_l$ and its second-harmonic $2\omega_l$ are expanded in the inset. The color-coded expanded inset of ITR-2PP shows interference fringes being tilted towards $t = 0$ fs with increasing final state energy (highlighted by the black line at ~24 fs). (c) 2D-FT photoelectron spectra obtained by Fourier transformation of the time axis in (a). The polarization energies induced in the sample at $0\omega_l$, $1\omega_l$ and $2\omega_l$ are correlated with the final state energies at which the photoemission process terminates. Top: Normalized Fourier filtered 2PP spectra at $0\omega_l$, $1\omega_l$ and $2\omega_l$. Right: line profiles taken at the SS and the IP state energies showing FT amplitudes of the $0\omega_l$, $1\omega_l$ and $2\omega_l$ oscillations.

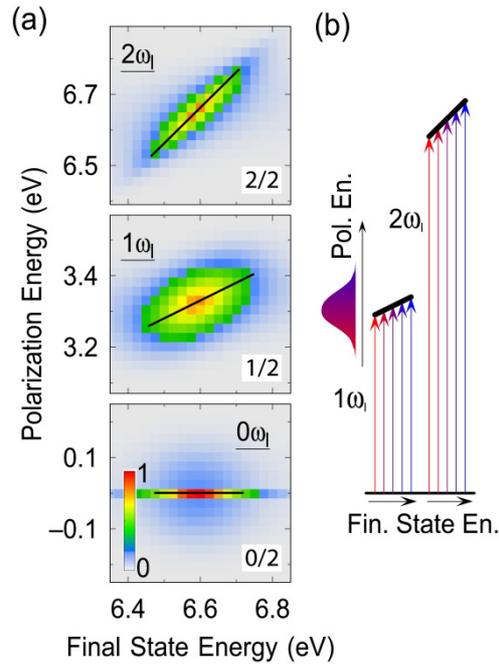

**Figure 4** | (Color online) (a) Expanded components of the OBE calculated 2D-FT spectra corresponding to the SS state signal from Fig. 3(b). The color table of each polarization order (0ω$_l$, 1ω$_l$, 2ω$_l$) is scaled separately. The 2D spectra for the zero-order component (0ω$_l$) is horizontal; the 2D spectra at the driving laser frequency (1ω$_l$) and its second-harmonic (2ω$_l$) disperse (tilt) with the final state energy; the slopes for a coherent process are indicated by black lines and quantified by the ratios in the white boxes. (b) Schematic illustration of the induced polarizations leading to the dispersive FT amplitudes in the 2D spectra: Each frequency component within the spectral range of the laser pulse terminates at a specific final state energy; left: linear polarizations (1ω$_l$), right: second-order polarizations (2ω$_l$).

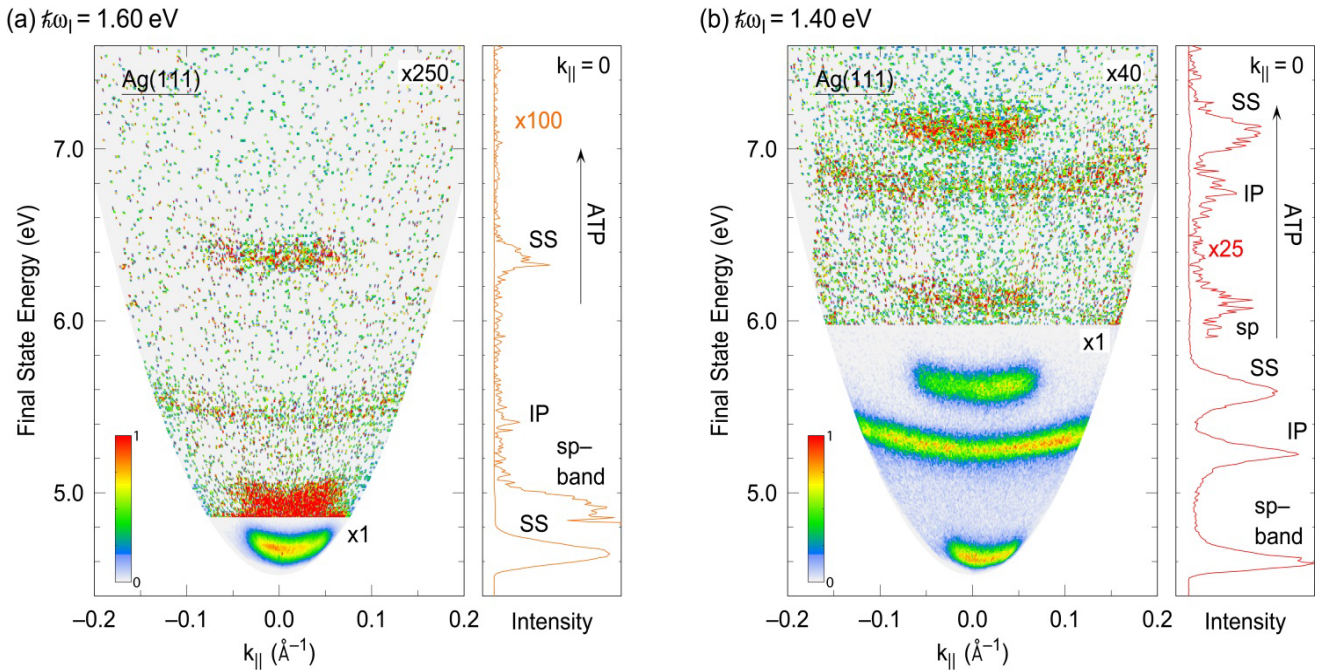

**Figure 5** | (Color online) Energy- and $k_\parallel$-resolved mPP spectra of Ag(111). Photon energies are chosen such that nonresonant excitation of the SS state is detected (a) by 3PP and (b) by 4PP; photoemission spectral features are labelled in the line profiles taken at $k_\parallel = 0$, excitation diagrams are given in Fig. 1(a). ATP contributions show replicas of states observed in 3PP and 4PP, but translated by one photon energy. (a) The photoemission spectrum excited with $\hbar\omega_l = 1.60$ eV has dominant contribution from the SS state excited by 3PP and weaker contributions from the $n = 1$ IP state and the direct sp-band transition, which are detected in 4PP. ATP feature attributed to $m = (3+1)$ excitation is found approximately $\approx 1.60$ eV above the 3PP SS state peak. (b) For $\hbar\omega_l = 1.40$ eV, the SS state, the $n = 1$ IP state, and the direct sp-band transition are detected in 4PP. At higher final state energies, their replicas appear in ATP through $m = (4+1)$ excitation.

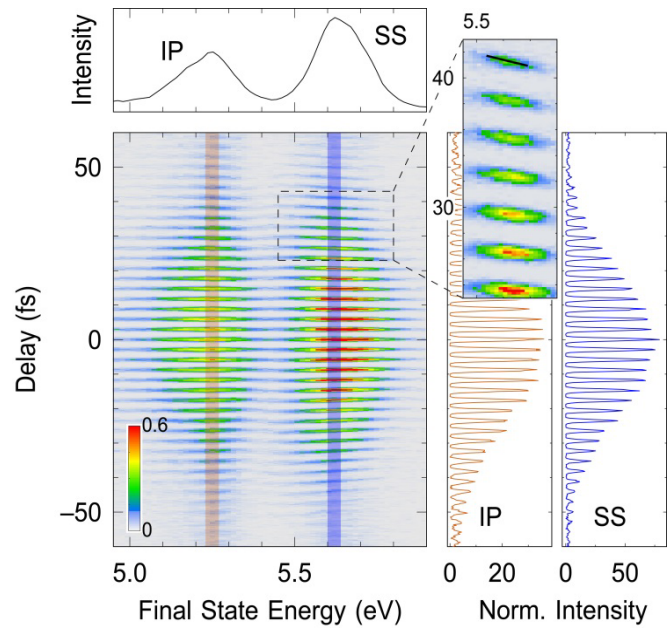

**Figure 6** | (Color online) ITR-4PP experiment on Ag(111) at a photon energy of $\hbar\omega_l = 1.40$ eV ($k_\parallel = 0$). Left bottom: Color-coded photoelectron counts as a function of pump-probe delay and final state energy; top: energy profile taken at $t = 0$ fs. Right: I2PC cross sections at the IP and SS final state energies. The high signal-to-baseline ratios are expected for the high-order correlation experiment. Inset: The interference fringes of the SS state tilt towards $t = 0$ fs with increasing final state energy; the tilt angle is marked by the black line at ~42 fs.

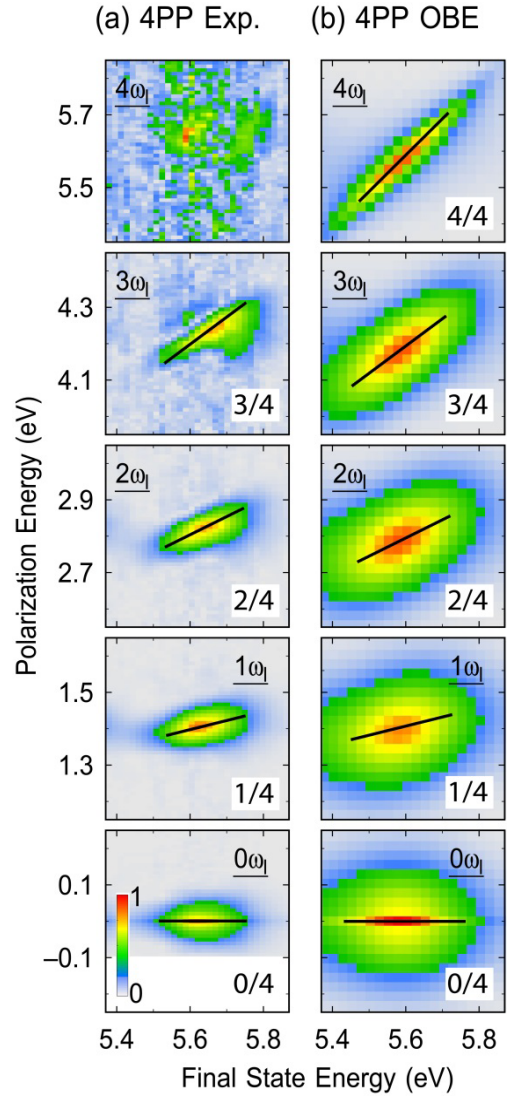

**Figure 7** | (Color online) 2D-FT photoelectron spectral components showing the SS state signal generated by Fourier analysis (a) of the ITR-4PP experiment in Fig. 6 and (b) a 5-level OBE simulation. The color-range of each 2D spectrum is scaled independently. FT amplitudes disperse with final state and polarization energy. The tilt is highlighted by black lines and their slopes are quantified in the white boxes.

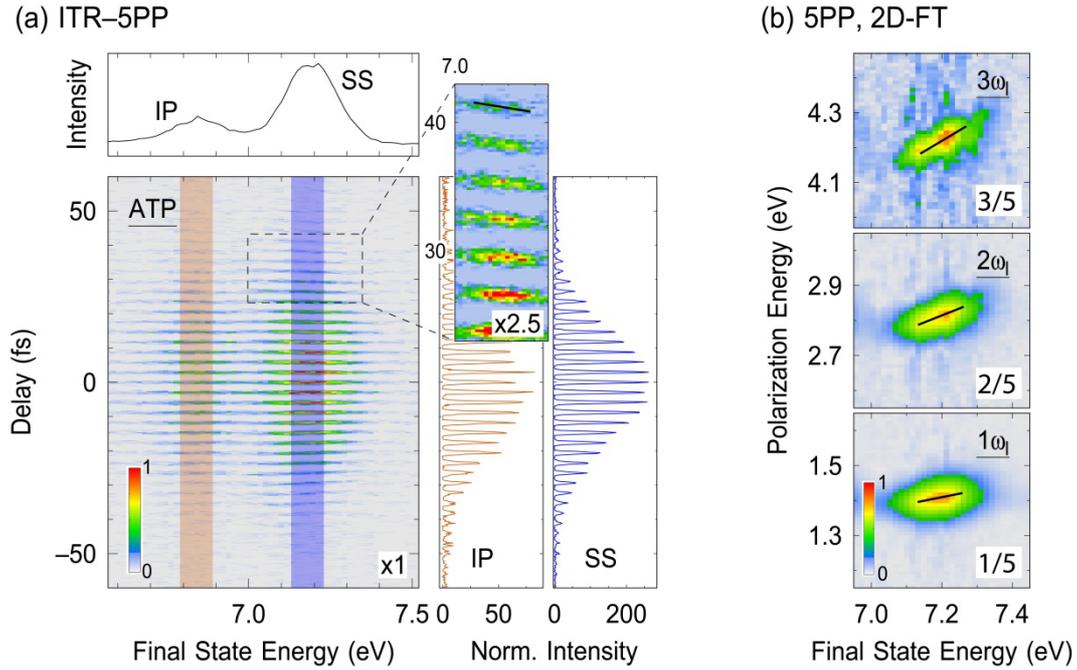

**Figure 8** | (Color online) (a) ITR-mPP experiment for the ATP component of the $n = 1$ IP and SS states that are excited in a $m = (4+1)$-photon process. The data is taken from the same experiment as shown in Fig. 6. Left bottom: Color-coded photoelectron counts as a function of pump-probe delay and final state energy. Top: energy profile taken for $t = 0$ fs. Right: The cross sections (I2PC scans) of the SS and IP states show a high peak-to-baseline ratio as expected for a $5^{th}$ order process. Inset: tilting of the fringes is highlighted by the black line at $t \approx 41$ fs. (b) Components of 2D-FT photoelectron spectra of the SS state excited in $m = (4+1)$-photon ATP process [FT of the data in (a)]. FT amplitude is observed for $0\omega_l$ (not shown), $1\omega_l$, $2\omega_l$ and $3\omega_l$; the slopes are marked with black lines and quantified in the white boxes. The color-range of each 2D spectrum is scaled independently.

# References.

# Coherent two-dimensional multiphoton photoelectron spectroscopy of metal surfaces

Marcel Reutzel[‡], Andi Li, and Hrvoje Petek[‡]

*Department of Physics and Astronomy and Pittsburgh Quantum Institute, University of Pittsburgh,*

*Pittsburgh, Pennsylvania 15260, USA*

*email: mar331@pitt.edu, petek@pitt.edu*

## S1. ITR-3PP

In Fig. S1, we show the 2D-FT photoelectron spectra of the SS state as obtained by Fourier transformation of an ITR-3PP experiment using $\hbar\omega_l = 1.60$ eV ($k_\parallel = 0$). The experiments are performed with the same parameters as used in the spectroscopy data of Fig. 5(a) in the main text. Along the interpretation of the ITR-4PP data in the main text, we fit the tilted FT components in the 2D-FT spectra with the slope $n/m$ and confirm thus the systematic dependence of the induced polarization signal ($n \leq m$) on the order of multi-photon excitation process ($m = 3, 4$). Our OBE calculations in a 4-level system confirm the extracted slopes [Fig. S1(b)].

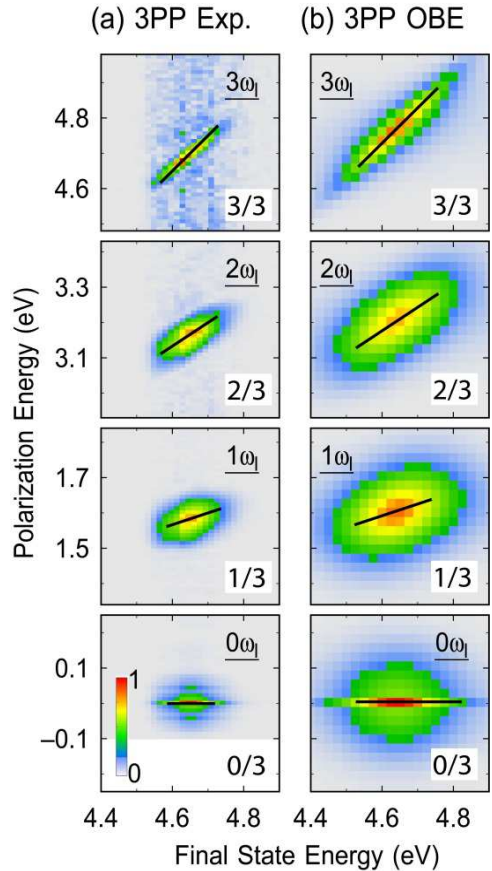

**Figure S1 |** 2D-FT photoelectron spectral components showing the SS signal generated by Fourier analysis (a) of the ITR-3PP experiment in Fig. 6 (main text) and (b) a 4-level OBE simulation. The color-range of each 2D spectrum is scaled independently. FT amplitudes disperse with final state and polarization energy. The tilt is highlighted by black lines and their slopes are quantified in the white boxes.